\documentclass[prl,twocolumn,tightenlines,superscriptaddress,preprintnumbers,floatfix]{revtex4}

\usepackage{graphicx}

\newcommand{\beq}{\begin{equation}}
\newcommand{\eeq}{\end{equation}}
\newcommand{\bqa}{\begin{eqnarray}}
\newcommand{\eqa}{\end{eqnarray}}
\newcommand{\msbar}{{\overline{\mbox{\rm MS}}}}
\newcommand{\LLH}{\chi^2}

\begin{document}

\title{Looking inside neutron stars: Microscopic calculations confront observations}

\preprint{\hskip5.5in\vbox{BI-TP 2010/17 \\ INT PUB 10-025}}

\author{Aleksi Kurkela}
\affiliation{Institute for Theoretical Physics, ETH Zurich, CH-8093 Zurich, Switzerland}
\author{Paul Romatschke}
\affiliation{Institute for Nuclear Theory, University of Washington, Box 351550, Seattle WA, 98195, USA}
\affiliation{Frankfurt Institute for Advanced Studies, D-60438 Frankfurt, Germany}
\author{Aleksi Vuorinen}
\affiliation{Faculty of Physics, University of Bielefeld, D-33501 Bielefeld, Germany}
\author{Bin Wu}
\affiliation{Faculty of Physics, University of Bielefeld, D-33501 Bielefeld, Germany}
\affiliation{School of Physics and State Key Laboratory of Nuclear Physics and Technology, Peking University, Beijing 100871, China}

\date{\today}

\begin{abstract}
While QCD appears not to be accurately solvable in the regime of interest for neutron star physics, microscopic calculations are feasible at both low and very high densities. In this work, we propose using the most realistic calculations in these two regimes of nuclear physics and perturbative QCD, and construct equations of state by matching the results requiring thermodynamic consistency. We find that the resulting equations of state --- in contrast to several hadronic ones --- are able to reproduce current observational data on neutron stars without any fine tuning, and allow stable hybrid stars with masses up to $2.1\,M_{\odot}$. Using recent observations of star radii, we perform a maximum likelihood analysis to further constrain the equation of state, and in addition show that the effects of rotation on radii and masses should not be neglected in future precision studies.
\end{abstract}

\maketitle

\paragraph{Introduction}

What are the thermodynamic properties of matter found in the center of a pulsar? Despite several decades of research in nuclear physics and a wealth of astrophysical data, this seemingly simple question has yet to be answered convincingly. It is well known that pulsars are made out of dense nuclear matter, and that the relevant interactions are described by Quantum Chromodynamics (QCD), but --- unlike at high temperature and low density --- accurately solving for the Equation of State (EoS) at high densities has proven to be impossible up to the present day. The situation is further complicated by the rich phase structure of cold nuclear matter, with possible phases containing \textit{e.g.~}kaon/pion condensation, hyperons and crystalline color superconductors \cite{Kaplan:1986yq,Schulze:2006vw,Alford:2007xm}. For these reasons, a commonly used approach in the study of high density nuclear matter is to give up on first principles calculations and resort to phenomenological models, such as those of the Nambu-Jona-Lasinio type (see \textit{e.g.~}Ref.~\cite{Buballa:2003qv} and references therein).

In this article, we propose a different strategy, based on the fact that accurate microscopic calculations for bulk thermodynamic quantities \emph{can} be performed for densities up to the nuclear saturation density $n_{\rm sat}=0.16\ {\rm fm}^{-3}$ in the hadronic phase \cite{Akmal:1998cf,Schulze:2006vw,Glendenning:1998zx} and at very high densities $n\gg n_{\rm sat}$ in the quark-gluon phase \cite{Kurkela:2009gj}. It is known that for sufficiently high density, a deconfinement transition from the hadronic to the quark-gluon phase must take place \footnote{We have chosen not to include a study of the scenario of absolutely stable strange quark matter in this work.}, and if one is interested in bulk thermodynamic quantities such as the EoS --- in contrast to the detailed phase structure of nuclear matter --- then it is plausible that quantitatively reliable results can be obtained by simply merging the high and low density results and requiring thermodynamic stability. Such an approach has indeed been found to be successful for nuclear matter at high temperature and low density, where a comparison to lattice QCD results is possible \cite{Kurkela:2009gj}.

In our work, we follow the general strategy outlined in Ref.~\cite{Kurkela:2009gj}, generalizing its results to be applicable for neutron stars that rotate and/or contain a mixed phase of nuclear and quark matter. While attempts to constrain the nuclear matter EoS based on astrophysical data have recently been carried out \cite{Ozel:2010fw,Steiner:2010fz}, our work constitutes the first direct comparison of state-of-the-art theoretical predictions with observations of rotating neutron stars.

\paragraph{Methodology}

We begin by providing a brief account of how candidate EoSs for neutron stars can be obtained from microscopic calculations. Readers interested in the details of the setup are kindly asked to consult Ref.~\cite{Kurkela:2009gj}.

First, we must identify the most reliable and accurate results available for the hadronic and quark matter EoSs. The primary uncertainty in the hadronic sector comes from the composition of the matter, which could include nucleons \cite{Akmal:1998cf}, nucleons and hyperons \cite{Schulze:2006vw}, or nucleons with a kaon condensate \cite{Glendenning:1998zx}. Less significant uncertainties enter \textit{e.g.~}from neglecting the interactions between more than three nuclei, the form of the variational ansatz, the unknown form of the condensation potential, as well as from the hyperon interactions, which we fix to the values considered realistic by the authors of Refs.~\cite{Glendenning:1998zx,Schulze:2006vw}. Our conclusion is that to obtain a conservative estimate for the uncertainty in the hadronic phase, it suffices to consider the three EoSs of Refs.~\cite{Akmal:1998cf,Glendenning:1998zx,Schulze:2006vw} as the limiting cases. For the very lowest densities, $n<n_{\rm{sat}}/2$, we use results from Refs.~\cite{Negele:1971vb,Baym:1971pw}.

In the description of the quark-gluon phase, it has been customary to use the MIT bag model \cite{Chodos:1974pn}, which provides an analytic EoS based on non-interacting massless quarks that is furthermore simple to parametrize. Whenever aiming for quantitative results, we, however, believe that it is not a fruitful approach to merge accurate (and complicated) calculations in the hadronic phase to the most simple-minded result in the quark-gluon phase, the popularity of which is to a large extent based on its ease of use. Thus, rather than using a model that altogether neglects quark interactions, we employ the most accurate perturbative QCD calculation available, which includes interactions of quarks and gluons to three loop order as well as a running strange quark mass \cite{Kurkela:2009gj}. The primary uncertainties that remain in the EoS are then the dependence of the result on the ratio of the renormalization scale to the baryon chemical potential $\bar{\Lambda}/\mu_B$ (which nevertheless vanishes order by order in perturbation theory) and the integration constant $B$, which is to some extent analogous to the bag constant of the MIT bag model.

Lesser uncertainties originate from the value of the strange quark mass $m_s$, the scale parameter of the $\msbar$ scheme $\Lambda_{\tiny \msbar}$ as well as the gap parameter $\Delta$, which enters the calculation should the quark-gluon matter be in a color superconducting phase. We have chosen to vary $3\bar{\Lambda}/\mu_B$ around its canonical value $2$ by a factor of two, and let $B$ take any value that is not in conflict with the positivity of the energy density or thermodynamic stability (\textit{cf.~}Ref.~\cite{Kurkela:2009gj}). The parameter $m_s$ is fixed using the most recent lattice results indicating $m_s(\bar{\Lambda}=2\,{\rm GeV})=92.2\pm1.3$ MeV \cite{McNeile:2010ji}, while $\Lambda_{\tiny \msbar}$ and $\Delta$ are let to vary in their expected ranges $346$ MeV $<\Lambda_{\tiny \msbar}<412$ MeV \cite{Kurkela:2009gj} and $0<\Delta<100$ MeV \cite{Alford:2007xm}. We stress that with $\Delta=100$ MeV, we probe the maximal effect that color superconductivity can have on the EoS.


\begin{figure}[t]
\center
\includegraphics[width=.8\linewidth]{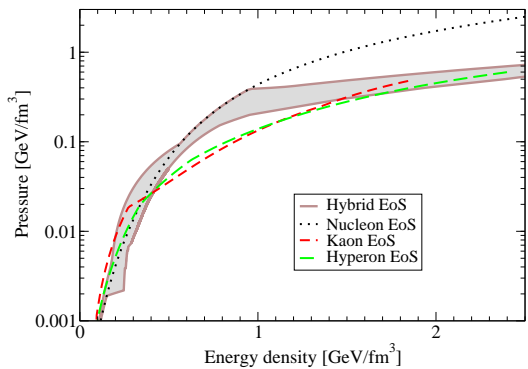}
\caption{Various EoSs resulting from the matching of the hadronic and quark matter phases. Also shown are the pure hadronic EoSs for nucleons, nucleons+hyperons and nucleons+condensed kaons. Requiring hadron and quark matter phases to match in a thermodynamically stable manner significantly reduces the range of the allowed EoSs.}
\label{fig:eos}
\vspace*{-0.5cm}
\end{figure}


While both the hadronic and quark matter calculations come with considerable uncertainty bands, it turns out that when requiring thermodynamically stable matching between the two phases, the uncertainty is significantly reduced \cite{Kurkela:2009gj} (see also Fig.~\ref{fig:eos}). We find that this is true both for homogeneous phases and when allowing a mixed phase of quarks and hadrons, the latter scenario being viable if the microscopic surface tension of QCD is low enough \cite{Alford:2001zr}. In both cases, we were unable to match the quark phase onto hadronic EoSs where hyperons are present or kaons have started to condense, because the pressure and energy density of these EoSs exceed those of quark matter at the same chemical potentials. Considering that at \emph{extremely} high densities, nuclear matter must be in the quark-gluon phase, we take this as an indication that neither hyperons nor kaon condensation are realized in cold nuclear matter. As will be demonstrated below, the same conclusion can be independently reached based on neutron star observations.

\paragraph{Results}

Neutron star observations have progressed significantly since the first discovery of a neutron star more than 40 years ago. The spin frequencies of more than 1800 objects have been catalogued to date \cite{ATNF}, the masses of approximately 70 neutron stars are known with some precision \cite{Stairs2004,Lattimer:2006xb}, and very recently, observational constraints on neutron star radii have become available \cite{Drake:2002bj,Poutanen:2003yd,Ozel:2010fw,Suleimanov:2010th} \footnote{Reported radii from Ref.~\cite{Ozel:2010fw} are in conflict with observations from Ref.~\cite{Suleimanov:2010th}. We are unable to resolve this conflict, but note that a reanalysis of the data of Ref.~\cite{Ozel:2010fw} in Ref.~\cite{Steiner:2010fz} is consistent with Ref.~\cite{Suleimanov:2010th} within error bars. Here, we have opted to use the results from Ref.~\cite{Steiner:2010fz}.}. In the following, we will use these observations in an attempt to constrain the cold nuclear matter EoS. Neutron stars with millisecond periods, for which accurate mass determinations exist, will turn out to be particularly useful in this process, and to this end we have collected all 15 presently known objects in Tab.~\ref{tab:one}.

\begin{table}[t]
\centering
\begin{tabular}{lccc}
Name & Spin (Hz) & Mass/${\rm M}_{\odot}$ & Remarks\\
\hline
J0024-7204H & 312 & $1.41\pm0.08$ &in 47 Tuc, \cite{Stairs2004}\\
J0437-4715  & 174 & $1.76\pm0.20$ &\cite{Verbiest:2008gy,ATNF}\\
J0514-4002A & 126 & $<1.52$       &in NGC1851, \cite{Freire:2007xg}\\
J0751+1807  & 288 & $1.26\pm0.14$ &\cite{Freire:2007xg}\\
J1012+5307  & 190 & $1.64\pm0.22$ &\cite{Stairs2004}\\
J1713+0747  & 219 & $1.53\pm0.08$ &\cite{Lattimer:2006xb,ATNF}\\
4U1608-52   & 619 & $1.70\pm0.40$ &LMXB, \cite{Ozel:2010fw,Deepto}\\
J1748-2446I & 105 & $1.85\pm0.05$ &in Ter 5,\cite{Freire:2007xg}\\
SAXJ1808.4-3658 &401 & $1.40\pm0.20$ &LMXB, \cite{Ozel:2008su}\\
J1824-2452C & 240 & $<1.37$       &in M28, \cite{Freire:2007xg}\\
B1855+09    & 187 & $1.58\pm0.13$ &\cite{Stairs2004}\\
J1903+0327  & 465 & $1.67\pm0.01$ &\cite{Champion:2008ge,ATNF}\\
J1909-3744  & 339 & $1.44\pm0.02$ &\cite{Lattimer:2006xb,ATNF}\\
J1911-5958A & 306 & $1.40\pm0.16$ &in NGC 6752,\cite{Freire:2007xg}\\
J2019+2425  & 254  &$<1.51$ &\cite{Stairs2004}\\
\end{tabular}
\caption{Masses of neutron stars with millisecond periods; see Refs.~\cite{Stairs2004,ATNF,Verbiest:2008gy,Freire:2007xg,Lattimer:2006xb,Ozel:2010fw,Deepto,Ferdman:2010rk,Champion:2008ge,Ozel:2008su} and references therein.}
\label{tab:one}
\vspace*{-0.5cm}
\end{table}

\begin{figure}[t]
\center
\includegraphics[width=.6\linewidth,angle=-90]{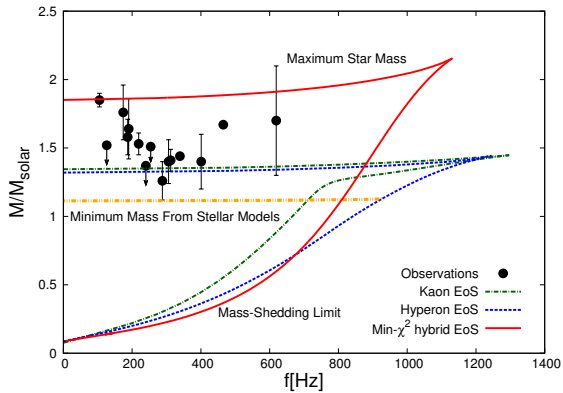}
\caption{Theoretical upper and lower limits for the masses of rotating neutron stars plotted against observational data. If the observed stars are approximately uniformly rotating, EoSs with hyperons and condensed kaons can be ruled out.}
\label{fig:mf_plot}
\end{figure}

While it is customary to use only mass-radius data to constrain the cold nuclear matter EoS, we emphasize that observations on neutron star masses and frequencies offer an interesting alternative because of the relatively accurate data available for many such systems (see Tab.~\ref{tab:one}). This is illustrated in Fig.~\ref{fig:mf_plot}, where the minimal and maximal neutron star masses allowed by the solutions to the general relativistic field equations for uniformly rotating stars are shown for a class of different EoSs. From here, it is seen that the EoSs containing hyperons and condensed kaons imply maximal masses for the stars that are significantly below the masses of several observed objects. This strengthens the above finding that hyperons and condensed kaons are most likely not realized in cold nuclear matter.

For EoSs consistent with observational data on millisecond period stars, we point out that stability with respect to radial oscillations as well as mass-shedding impose further limits on the masses, and that this range is rapidly shrinking for the fastest spinning objects. The maximum masses tend to increase markedly with the spin frequency $f$, for some EoSs as much as $20$ percent in comparison with $f=0$. On the other hand, using the theoretical minimum mass for the iron core of a progenitor star as input \cite{Woosley:2007as}, current supernova simulations suggest that newborn neutron stars have a minimum (gravitational) mass of $\sim 1.1 M_{\odot}$ \cite{Huedepohl:2009wh} (also indicated in Fig.~\ref{fig:mf_plot}). As a consequence of the frequency dependence of these lower and upper limits, we expect stars with higher spin rates to have higher masses on average. Improvements in our understanding of core-collapse supernovae together with future observations could make it possible to eventually accentuate the conclusions to be drawn from Fig.~\ref{fig:mf_plot}.

\begin{table}[t]
\centering
\begin{tabular}{lcc}
Name & Mass/${\rm M}_{\odot}$ & Radius/km\\
\hline
4U1608-52   & $1.70\pm0.43$ & $11.5\pm2.0$\\
4U1742-307  & $1.63\pm0.58$ & $14.4\pm1.0$\\
EXO1745-248 & $1.55\pm0.40$ & $10.0\pm1.8$\\
SAXJ1808.4-3658 & $1.40\pm0.20$ &$10.0\pm4.0$\\
4U1820-30   & $1.73\pm0.33$   & $11.5\pm1.5$\\
RXJ1856-375 & $1.70\pm0.30$     & $11.5\pm1.2$
\end{tabular}
\caption{Known neutron star masses and radii \cite{Poutanen:2003yd,Ozel:2010fw,Steiner:2010fz,Suleimanov:2010th}.}
\label{tab:two}
\vspace*{-0.7cm}
\end{table}

\begin{figure}[t]
\center
\includegraphics[width=.8\linewidth]{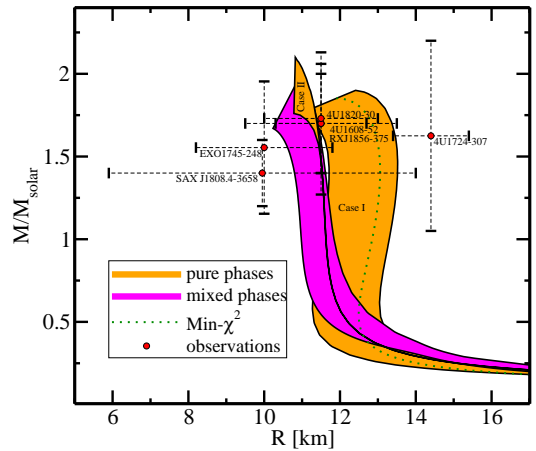}
\caption{Star mass-radius curves for hybrid EoSs compared to observations (both assuming negligible spin frequencies). The pure phases region is split into two parts according to the `Case I' and `Case II' matching criteria of Ref.~\cite{Kurkela:2009gj}.}
\label{fig:MR}
\vspace*{-0.7cm}
\end{figure}

For those EoSs that have not been ruled out above, it is possible to continue the comparison with available mass-radius data. At present, neutron star radii have been determined with reasonable accuracy for altogether 6 stars (Tab.~\ref{tab:two}), so that for two of them, 4U1608-52 and SAXJ1808.4-3658, information on three different quantities --- mass, radius and spin --- is available. Neutron star radii are customarily determined assuming the stars to be non-rotating. If we follow this assumption and perform a maximum likelihood analysis of the six objects in Tab.~\ref{tab:two}, we find that EoSs generating stars with a large quark core and only a small hadronic crust are favored. The likelihood is measured in terms of a merit function
$
\LLH \equiv \sum_{i=1}^6 {\rm Min}[(M-M_i)^2/\sigma_{Mi}^2+(R-R_i)^2/\sigma_{Ri}^2],
$
where $M_i,\,R_i$ are the masses and radii from Tab.~\ref{tab:two}, $\sigma_{Mi}$ and $\sigma_{Ri}$ are the respective standard deviations, and the minimization is performed along the $MR$ curve generated by the EoS. We observe that the large central value and small error bars for the radius of 4U1724-307 have the effect of disfavoring EoSs with small radii (\textit{e.g.~}those with mixed phases) and favoring ones where the phase transition takes place at moderately small densities.

Overall, all of the hybrid EoSs that we obtain by matching the hadron and quark phases result in mass-radius curves that are reasonably close to observations, as can be seen in Fig.~\ref{fig:MR}. For slowly rotating hybrid stars, masses up to 2.1 $M_{\odot}$ are possible, but these stars contain only a tiny quark core at the center of the star. When considering a mixed phase of hadrons and quarks, the maximal masses encountered are 1.93 $M_{\odot}$, and for hybrid stars with only a small hadronic crust, we find masses less than 1.9 $M_{\odot}$. In general, for hybrid stars with masses of $1.5$ $M_\odot$, the radii range from 10.8 to 13.5 km, with 13.1 km corresponding to the smallest $\LLH$ value when attempting to fit to observations (\textit{cf.~}Fig.~\ref{fig:MR}).

\begin{table}[t]
\centering
\begin{tabular}{ccccc}
$\Delta/ {\rm MeV}$ &
$3\bar{\Lambda}/(\mu_B\Lambda_{\tiny \msbar})\times{\rm GeV}$&
$\mu_{{\rm match}}/{\rm MeV}$&
mixed phase &
$\LLH$\\
\hline
100 & 9.26  &  955 & no &4.42 \\ 
0     & 11.56 & 960 & no & 5.93 \\ 
100 & 11.56 & 1430 & yes & 9.07 \\ 
0    &  8.67  & 1617 & yes & 9.16 \\ 
100 & 10.58 &  1222 & no & 9.70 \\ 
0     & 11.56 & 1492 & no & 10.86 \\ 
0     & 11.56 & 1492 & yes & 10.87 \\ 
100 & 11.56 & 	1257 & yes & 14.48	\\ 
\end{tabular}
\caption{$\LLH$ values for selected hybrid EoSs with quark matter (parameters $\Delta, 3\bar{\Lambda}/\mu_B$ and $\Lambda_{\tiny \msbar}$) matched to nucleons at $\mu_B=\mu_{{\rm match}}$. The set consists of the best and worst fits for $\Delta=0,0.1$ GeV and with and without mixed phases. For the purely nucleonic equation of state, we obtain $\LLH=9.16$.}
\label{tab:three}
\end{table}

\begin{figure}[t]
\center
\includegraphics[width=.5\linewidth,angle=-90]{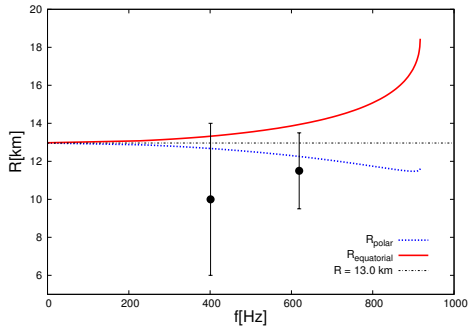}
\caption{Observational data for the radii and frequencies of the stars SAXJ1808.4-3658 and 4U1608-52. Using the minimum $\chi^2$ hybrid EoS ($\Delta=0.1$ GeV, $3\bar{\Lambda}/(\mu_B\Lambda_{\tiny \msbar})=11.56$ ${\rm GeV}^{-1}$, $\mu_{{\rm match}}=956$ MeV, no mixed phase) and assuming a star mass of $M/M_{\odot}\simeq 1.5$, results for both the polar and equatorial radii as functions of frequency are shown for comparison.}
\label{fig:rfplot}
\vspace*{-0.5cm}
\end{figure}

Finally, we note that the assumption of non-rotation in the analysis of neutron star radii introduces systematic errors, as many of the stars are in fact rapidly rotating. In Fig.~\ref{fig:rfplot}, we demonstrate how the polar and equatorial radii of a star with a constant baryonic mass start to differ strongly at large frequencies. When aiming at sub-km uncertainties in radii observations and theory analysis in the future, the effect of rotation will thus have to be taken into account for frequencies exceeding $600$ Hz.

\paragraph{Conclusions}

We have constructed possible equations of state for neutron star interiors by matching state-of-the-art microscopic calculations at low and high densities (hadronic and quark matter phases, respectively). Our findings show that these EoSs lead to neutron star masses and radii that are consistent with present observations without having to fine tune any parameters. Using current observational data, we are able to rule out hadronic equations of state that contain either hyperons or assume kaon condensation. A maximum likelihood analysis seems to favor neutron stars consisting of a large quark matter core with only a thin hadronic crust, but --- given the scarcity of objects, for which sufficient data is available --- this conclusion must be considered tentative. Finally, we have pointed out how to further constrain the EoS once additional observational information on neutron star masses, frequencies and radii becomes available, in particular for rapidly rotating systems.

All EoSs used in this article are available online \cite{web}.

\paragraph{Acknowledgements}
We would like to thank D.~Chakrabarty, T.~Janka, F.~Ozel, J.~Poutanen, B.~Rutledge, J.~Schaffner-Bielich, H.~J.~Schulze, A.~Steiner and S.~Woosley for fruitful discussions. The work of AK was supported by the SNF grant 20-122117, while PR was supported partially by the US Department of Energy, grant number DE-FG02-00ER41132, and partially by the Helmholtz International Center for FAIR within the framework of the LOEWE program launched by the State of Hesse. AV and BW were supported by the Humboldt foundation through its Sofja Kovalevskaja program. The computations for rotating stars were performed using the code of Ref.~\cite{Stergioulas:1994ea} (http://www.gravity.phys.uwm.edu/rns/).

\end{document}